\documentclass[aps,prb,floatfix,twocolumn,nofootinbib]{revtex4-1}

\usepackage{amsmath,amsfonts,amssymb}
\usepackage{graphicx}

\begin{document}

\title{Broadband parametric amplification with impedance engineering: Beyond the gain-bandwidth product}

\author{Tanay Roy$^{1}$, Suman Kundu$^{1}$, Madhavi Chand$^{1}$}
\author{A. M. Vadiraj$^{1}$}
\altaffiliation{Current address: Institute for Quantum Computing \& Electrical and Computer Engineering Department, University of Waterloo, Waterloo N2L 3G1, Canada.}
\author{ A. Ranadive$^{1}$, N. Nehra$^{1}$,  Meghan P. Patankar$^{1}$, J. Aumentado$^{2}$, A. A. Clerk$^{3}$}
\author{R. Vijay$^{1}$}
\email{r.vijay@tifr.res.in}

\affiliation{$^{1}$Department of Condensed Matter Physics and Materials Science, Tata Institute of Fundamental Research, Homi Bhabha Road, Mumbai 400005, India}
\affiliation{$^{2}$National Institute of Standards and Technology, Boulder, Colorado 80305}
\affiliation{$^{3}$Department of Physics, McGill University, 3600 rue University, Montr{\'e}al, Quebec H3A 2T8, Canada}


\date{\today}

\begin{abstract}
We present an impedance engineered Josephson parametric amplifier capable of providing bandwidth beyond the traditional gain-bandwidth product.  We achieve this by introducing a positive linear slope in the imaginary component of the input impedance seen by the Josephson oscillator using a $\lambda/2$ transformer. Our theoretical model predicts an extremely flat gain profile with a bandwidth enhancement proportional to the square root of amplitude gain. We experimentally demonstrate a nearly flat 20 dB gain over a 640 MHz band, along with a mean 1-dB compression point of -110 dBm and near quantum-limited noise. The results are in good agreement with our theoretical model.
\end{abstract}

\maketitle

Josephson parametric amplifiers (JPAs) have become a crucial component of superconducting qubit \cite{devoret_science_rev} measurement circuitry, enabling recent studies of quantum jumps \cite{quantum-jump-vijay}, generation and detection of squeezed microwave field \cite{state-tomo}, quantum feedback \cite{rabi-stabilize-vijay, dicarlo-fb}, real-time tracking of qubit state evolution \cite{hatridge-meas-backaction, qtraj-murch,optimal-route-weber} and quantum error detection \cite{err-detect-IBM, bit-flip-riste}. Although JPAs based on Josephson junctions embedded in a resonator \cite{magnetometry-hatridge, lehnert-paramp, JPC-amp} regularly achieve 20 dB power gain and quantum-limited noise, typical bandwidth is restricted to 10-50 MHz \cite{magnetometry-hatridge, LJPA-mutus}, making them suitable for single qubit measurements only. The rapid progress towards multi-qubit architectures \cite{err-detect-IBM} for fault-tolerant quantum computing \cite{multiplex-martinis, surface-code-fowler} demands an amplifier with much larger bandwidth to enable simultaneous readout of multiple qubits with minimal resources.

There have been several attempts in this direction in recent years. One such attempt used a broadband impedance transformer \cite{impa} to lower the quality factor of a lumped-element Josephson oscillator which is the main component of a parametric amplifier. While the observed large bandwidth was qualitatively explained by a model consisting of a negative resistance \cite{pumpistor} coupled to a frequency dependent impedance, no clear prescription on the design principle was provided. A different approach using Josephson non-linear transmission lines \cite{TWPA-theory-1, TWPA-theory-2} was recently demonstrated \cite{TWPA-martinis,TWPA-Berkeley} with nearly 4 GHz of bandwidth. However, this design requires fabrication of about 2000 nearly identical blocks of oscillator stages, demanding fairly sophisticated fabrication facilities.
Multimode systems utilizing dissipative interactions have also been suggested theoretically as a route for enhancing bandwidth \cite{anja_prl}, but have not been realized experimentally.

In this Letter, we present a simple technique for enhancing the bandwidth of a JPA and beating the standard gain-bandwidth limit.  It involves engineering the {\it imaginary} part of the environmental impedance:  in particular, 
we introduce a positive linear slope in the imaginary component of the impedance shunting the JPA, while keeping the real part unchanged at the pump frequency. Our design uses a combination of a $\lambda /4$ and a $\lambda /2$ impedance transformers which are significantly easier to fabricate than a broadband impedance transformer \cite{impa}. Our theoretical model explains why the imaginary part of the impedance plays a crucial role in determining the amplifier bandwidth, and predicts an extremely flat gain profile with a bandwidth beyond the standard gain-bandwidth product. We experimentally demonstrate 640 MHz of nearly flat 3 dB bandwidth with 20 dB peak gain at a center frequency of $5.93$ GHz. The observed near-quantum-limited noise performance with a mean 1-dB compression point of -110 dBm over the whole bandwidth makes it a promising candidate for use in multi-qubit architecture.

We model the JPA as a non-linear resonator with a plasma frequency $\Omega_p$ coupled to a frequency dependent environment specified by an input impedance $Z_\text{in}[\omega]$. We consider the directly pumped (at $\Omega_d$) case \cite{magnetometry-hatridge} but our approach is equally applicable to flux pumping \cite{flux-driven-Yamamoto, impa} as well. Similar to Ref.~\onlinecite{impa}, we will be interested in using the JPA in a phase-preserving mode of operation.  It is thus useful to treat the signal and idler degrees of freedom of the single physical cavity mode as two separate modes \cite{supp}. At the mean-field level, the system is then analogous to a detuned non-degenerate parametric amplifier (NDPA) \cite{clerk_paramp,clerk-laflamme-PRA}.  In an interaction picture at the drive frequency, the effective Hamiltonian  in the rotating wave approximation is given by
\begin{equation}
	\label{eq:jpahamil}
	\hat{H}_A= \omega_{d}' \left( \hat{a}_1^\dagger\hat{a}_1 +  \hat{a}_2^\dagger\hat{a}_2 \right) 
		-i \lambda (\hat{a}_1^\dagger\hat{a}_2^\dagger-h.c.),
\end{equation}
where $\lambda$ is the strength of effective parametric pumping. Here, $\hat{a}_1$ and $\hat{a}_2$ are canonical lowering operators describing the signal and idler frequency components of the same physical cavity mode; they can be treated as independent as long as the signal frequency is not  equal to the pump frequency \cite{supp}.  
Also,  $\omega_d'=\omega_d-2\lambda$ where $\omega_d=\Omega_p-\Omega_d >0$ is the detuning of the pump from the oscillator resonant frequency. The effective detuning $\omega_d'$ depends on the pumping strength $\lambda$ for parametric amplifiers based on linear driving of a quartic Kerr non-linearity \cite{clerk-laflamme-PRA}; in contrast, if one flux-pumps, one has $\omega_d' = \omega_d$ (and could thus easily tune to have $\omega_d' = 0$).

Solving the equations of motion\cite{supp,dykman1994} yields the system susceptibility matrix $\chi[\omega]$ that relates the intracavity fields $\vec{a}[\omega]=(\hat{a}_1,\hat{a}_2^\dagger)$ to the input fields $\vec{a}_{\rm in}[\omega]$ as $\vec{a}[\omega]=\chi[\omega]\vec{a}_{\rm in}[\omega]$, where
\begin{equation}
\label{eq:genchi}
(\chi[\omega])^{-1}=
-i\begin{pmatrix}
(\omega-\omega_{d}')-\Sigma_{1}[\omega]& i\lambda \\
i\lambda& (\omega+\omega_{d}')-\Sigma_{2}[\omega]
\end{pmatrix}.
\end{equation}
Here $\omega$ is the signal frequency detuning from $\Omega_d$ while $\Sigma_{1,2}[\omega]$ is the self energy and is related to the input admittance ($Y_\text{in}[\omega] = 1/Z_\text{in}[\omega]$) seen by the JPA as \cite{supp}
\begin{subequations}
\label{eq:sigmaYin}
\begin{align}
\Sigma_{1}[\omega]&=\Delta_{1}[\omega]-i{\kappa}_{1}[\omega]/2 \ = - \frac{ i }{2 C_p}  Y_{\rm in}^* [\omega],  \\
\Sigma_{2}[\omega]&=\Delta_{2}[\omega]-i{\kappa}_{2}[\omega]/2 \ = - \frac{ i }{2 C_p}  Y_{\rm in} [-\omega],
\end{align}
\end{subequations}
where $\kappa_{1(2)}[\omega]$ and $\Delta_{1(2)}[\omega]$ are related to the real and imaginary parts of $Y_{\text{in}}[\omega]$ at the signal ($+\omega$) and idler ($-\omega$) frequencies respectively.
The photon number gain of the signal mode $\mathcal{G}[\omega]$ is given by 
\begin{subequations}
\label{eq:gain}
\begin{align}
\mathcal{G}[\omega]&=|{r}[\omega]|^2=|1-\kappa_1\chi_{11}[\omega]|^2, \\ 
\chi_{11}[\omega]&= \dfrac{-i(\omega+\omega_d'-\Delta_2[\omega])+\kappa_2[\omega]/2}{(-i\tilde\omega[\omega]+\kappa_-[\omega]/2)(-i\tilde\omega[\omega]+\kappa_+[\omega]/2)},
\end{align}
\end{subequations}
with
\begin{subequations}
\label{eq:omegakappa}
\begin{align}
\tilde\omega[\omega] &=\omega-(\Delta_1[\omega]+\Delta_2[\omega])/2, \\
 \kappa_\pm[\omega] &=(\kappa_1[\omega]+\kappa_2[\omega])/2\pm2\sqrt{(\lambda^2-\omega_d'^2)+\epsilon[\omega]},
\end{align}
\end{subequations}
Here $\epsilon[\omega]$ is a small quantity \cite{supp} which depends on the symmetry of $Y_\text{in}[\omega]$ about $\omega=0$. Both the peak gain $\mathcal{G}[0]$ and bandwidth (FWHM of $\mathcal{G}[\omega]$) are controlled by the ``slow" effective damping rate $\kappa_-[\omega]$.

\begin{figure}[t]
\includegraphics[width=0.5\textwidth, bb = 0 0 243.8 141.8]{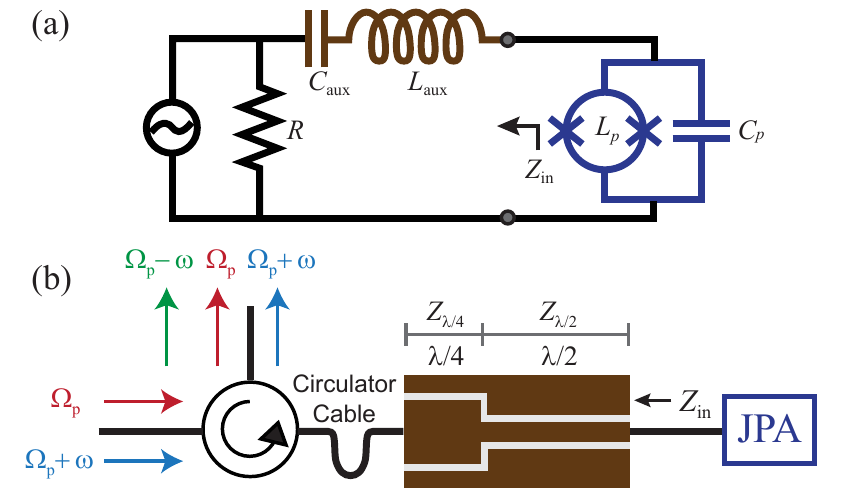} 
\caption{Impedance engineered parametric amplifier. (a) The non-linear resonator composed of a capacitively shunted SQUID is coupled to the source impedance via a series LC circuit which introduces a positive linear slope in the imaginary part of the input impedance $Z_\text{in}[\omega]$. (b) Implementation of the circuit in (a) using a $\lambda/2$ impedance transformer. The $\lambda/4$ section is used to transform the 50 $\Omega$ environment to about 30 $\Omega$.}
\label{fig:parampLC}
\end{figure}

For the simplest case of the JPA coupled to a frequency-independent environment with characteristic impedance $R$, $Y_\text{in}[\omega] = 1/R$, leading to $\kappa_{1(2)}[\omega]=\kappa_0$, $\Delta_{1(2)}[\omega]=0$ and $\epsilon[\omega]=0$. Here $\kappa_0=\Omega_p Z_p/R$ is the oscillator linewidth , with $Z_p=(L_p/C_p)^{1/2}$ being the characteristic impedance of the oscillator (Fig.~\ref{fig:parampLC}(a)). For a given $\omega_d$, there is an optimal $\lambda$ which maximizes $\mathcal{G}[0]$ and is given by \cite{magnetometry-hatridge}
\begin{equation}
\label{eq:gmax}
\mathcal{G}_\text{max}=1+\left(3\left(1+7\beta^2-4\beta\sqrt{1+3\beta^2}\right)\right)^{-1},
\end{equation}
where $\beta=2\omega_d/(\sqrt{3}\kappa_0) \lesssim 1$, while the gain and bandwidth (in the large gain limit) are given by
\begin{equation}
\label{eq:FWHM}
	\mathcal{G}[\omega] \approx \dfrac{\mathcal{G}_\text{max}}{1+ (\omega/\Gamma_\text{BW})^2}, \ \  
	2\Gamma_{\text{BW}} \approx 
		\kappa_0 \left( \dfrac{ 1 }{\mathcal{G}_\text{max}} \right )^{1/2}.
\end{equation}

Eq.~\eqref{eq:FWHM} gives the standard gain-bandwidth product \cite{clerk_paramp} where the bandwidth falls off as the square-root of power gain. For a given $\mathcal{G}_\text{max}$,  the simplest way to enhance bandwidth \cite{impa} is to increase $\kappa_0$ by reducing the shunt impedance $R$. However, lowering the quality factor $Q=\Omega_p/\kappa_0$ requires one to pump harder and at some point the strong pump makes the amplifier unstable and noisy due to higher order non-linearities \cite{vlad-bif} which are ignored in our linearized Hamiltonian (Eq.~\eqref{eq:jpahamil}). 

We consider a different approach where we leave the shunting impedance at the pump frequency ($\omega=0$) undisturbed so that the same optimal value of $\lambda$ provides $\mathcal{G}_\text{max}$ as before.  Instead of increasing $\kappa_0$, we modify the coherent cavity dynamics to make the amplifier less sensitive to the effects of detuning the signal from the pump (i.e.~non-zero $\omega$); as such, strong amplification will persist even away from $\omega = 0$. This can be achieved by introducing a frequency dependence in $Y_\text{in}[\omega]$ with the help of an auxiliary system such that $\Sigma_{1(2)}[\omega]$ exactly cancels the explicit $\omega$ dependence of $\chi[\omega]$ in Eq. \eqref{eq:genchi}. The result would be a modified susceptibility ${\tilde\chi}[\omega]$ and gain $\tilde{\mathcal{G}}[\omega]$ that are frequency independent.
A simple step in this direction is to use an additional linear resonant mode as the auxiliary system.  Here, we introduce a series LC resonator as shown in Fig.~\ref{fig:parampLC}(a). We further make the optimal choice of having the auxiliary LC resonance frequency match the drive frequency i.e.~$\Omega_\text{aux}=\Omega_d$. For small $\omega$, the input impedance can now be written as
\begin{equation}
\label{eq:zinideal}
Z_{\text{in}}[\omega]=R+i\alpha \omega,
\end{equation}
where $\alpha={2Z_{\rm aux}}/{\Omega_{\rm aux}}$ is the slope of the imaginary component and $Z_{\rm aux}=\sqrt{L_{\rm aux}/C_{\rm aux}}$ is the characteristic impedance of the  auxiliary resonator. This choice of $Z_\text{in}[\omega]$ leads to $Y_\text{in}[-\omega] = Y_\text{in}[\omega]^*$ and hence
\begin{subequations}
\label{eq:deltakappa}
\begin{align}
\Delta_{1(2)}[\omega]&=\dfrac{1}{2 \alpha C_p}\left(\dfrac{\omega}{\omega^2 + R^2/\alpha^2}\right), \\
{\kappa}_{1(2)}[\omega]&=\dfrac{R}{\alpha^2 C_p}\left(\dfrac{1}{\omega^2 + R^2/\alpha^2}\right),
\end{align}
\end{subequations}
with $\epsilon[\omega]=0$. For small $\omega$, $\Delta[\omega] \sim \alpha \omega/(2R^2C_p)$ and can be used to cancel the intrinsic $\omega$ dependence in $\chi[\omega]$ to leading order: one just tunes $\alpha$ by adjusting $Z_{\rm aux}$. We see that it is $\alpha$ which controls the frequency-dependence of the ``induced potential" $\Delta_{1(2)}[\omega]$ that is crucial to our scheme while $\kappa_{1(2)} \sim 1/(2RC_p) = \kappa_0$ is constant for small $\omega$. The modified gain can now be computed by using Eqs.~\eqref{eq:gain}, \eqref{eq:omegakappa} and \eqref{eq:deltakappa}.
\begin{figure}[t]
\includegraphics[width=0.5\textwidth, bb = 0 0 243.8 104.9]{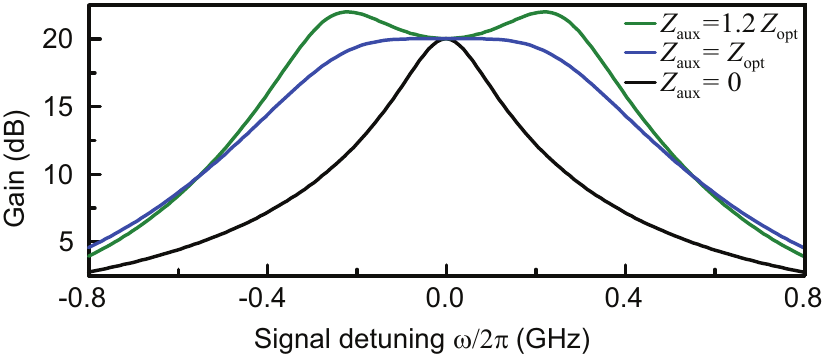}
\caption{Theoretical gain curves as a function of signal detuning from pump frequency for different values of $Z_{\rm aux}$. The gain is flattest when $Z_{\rm aux}=Z_\text{opt}$.}
\label{fig:thgain}
\end{figure}

The leading-order quadratic dependence of $\tilde{\mathcal{G}}[\omega]$ in our modified JPA can be canceled exactly for an optimal choice of $Z_{\rm aux}$, given by
\begin{equation}
\label{eq:optz0}
Z_\text{opt}=\eta R^2 / Z_p,
\end{equation}
where $\eta$ is a prefactor of order unity \cite{supp}. The zero frequency gain is still given by $\mathcal{G}_\text{max}$ since $Z_\text{in}[\omega=0]$ is left unmodified in our scheme
(i.e.~${\kappa}_{1(2)}[0] = \kappa_0, \Delta_{1(2)}[0]=0$), while the gain $\tilde{\mathcal{G}}[\omega]$ roll-off is much slower near $\omega=0$.  In the large gain limit, the modified gain profile and corresponding bandwidth are given by
\begin{equation}
\label{eq:newBW}
	\mathcal{\tilde{G}}[\omega]=\dfrac{\mathcal{G}_\text{max}}{1+ (\omega /\Gamma_\text{BW})^4}, \ \ 
	2\tilde\Gamma_{\text{BW}} \approx 
		\kappa_0 \left( \dfrac{ 1 }{\mathcal{G}_\text{max}} \right )^{1/4},
\end{equation}
with a bandwidth enhancement factor $\mathcal{F}= \tilde\Gamma_{\text{BW}} / \Gamma_{\text{BW}} \sim (\mathcal{G}_\text{max})^{1/4} $. For finite gain, $\mathcal{F}$ is somewhat larger than this asymptotic value. 

Fig.~\ref{fig:thgain} shows how the frequency-dependence of the gain is modified for different choices of  $Z_{\rm aux}$; the flattest profile is obtained for  $Z_{\rm aux}=Z_\text{opt}$. For a value of $Z_{\rm aux}$ larger than the optimal value, the gain profile transforms to a double peaked shape (green curve of Fig.~\ref{fig:thgain}) where the two peaks correspond to the normal modes of the two coupled oscillators. The optimal value for $Z_{\rm aux}$ ensures that the contribution of the two modes overlap in just the right way to give a flat gain response near $\omega=0$. 
\begin{figure}[t]
\includegraphics[width=0.5\textwidth, bb = 0 0 244 171.7]{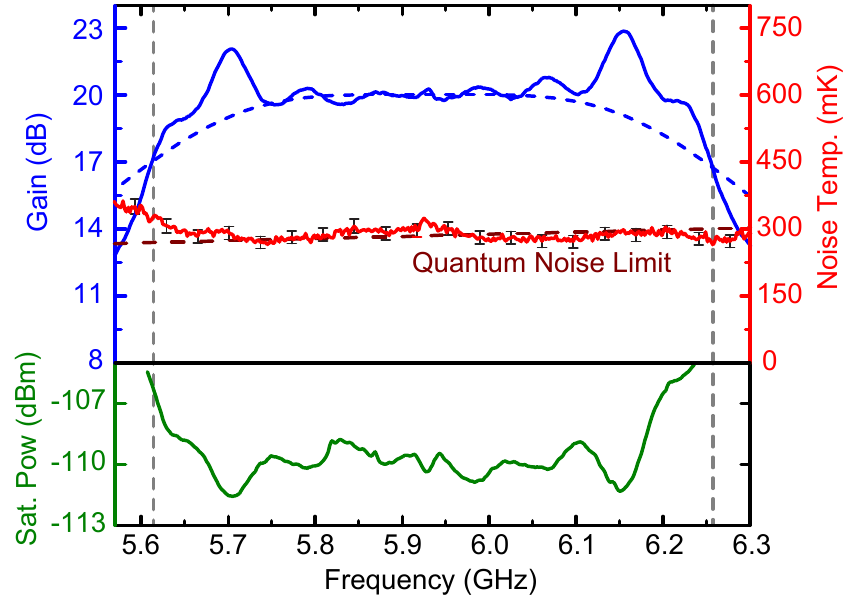} 
\caption{Experimentally measured gain (solid blue), noise temperature (solid red) and 1-dB compression point (solid green) as a function of signal frequency. The gain at the pump frequency of 5.93 GHz is 20 dB. The gain profile is mostly flat with a few ripples and shows good agreement with the theoretical prediction (dashed blue). The noise is near quantum limited over the 640 MHz (-3 dB) bandwidth indicated by the vertical dashed lines.}
\label{fig:gain}
\end{figure}

Having established the basic theory underlying our design, we now discuss its experimental implementation.  This was done in two steps (Fig.~\ref{fig:parampLC}(b)). First, we used a $\lambda/4$ section (at $\Omega_d$) to transform the standard $R=50 \ \Omega$ environment to $R=$ Re$[Z_\text{in}[0]] = 30 \ \Omega$. This not only provides some conventional bandwidth enhancement but also keeps the value of $Z_\text{opt}\sim 80 \ \Omega$ (for $Z_p \sim 6.5 \ \Omega$) in a range which is easy to implement. Second, we used a $\lambda/2$ section (at $\Omega_d$) to independently set the slope of Im$[Z_\text{in}[\omega]]$, and hence $Z_\text{opt}$, while keeping $R$ unchanged. Both transformer sections were made on a single RF circuit board in the co-planar waveguide (CPW) geometry for a design frequency of 6 GHz. 

Our JPA device was made using standard e-beam lithography and double-angle evaporation to fabricate \cite{supp} the SQUID with a total critical current $\sim 2.8$ $\mu$A and the capacitor $C_p\sim3.4$ pF. The SQUID geometry allowed the tuning of $\Omega_p / 2\pi$ in the range 5 - 7.25 GHz with the help of a superconducting coil. The device was connected to the impedance transformer using wire-bonds and measured at 20 mK temperature inside a dilution refrigerator. We used standard cryogenic microwave reflectometry setup to measure the devices \cite{supp}. A magnetically shielded cryogenic circulator connected to the JPA was used to separate the incident and reflected signal and both were housed inside a magnetic and radiation shielded can.

\begin{figure}[t]
\includegraphics[width=0.5\textwidth, bb = 0 0 243.8 174.7]{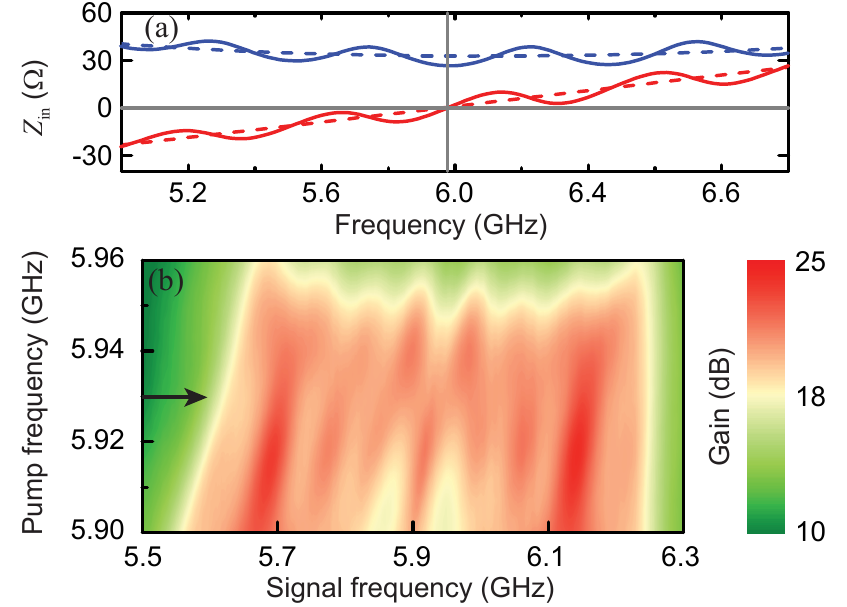} 
\caption{(a) Real (blue) and imaginary (red) parts of the input impedance $Z_{\rm in}[\omega]$ seen by the parametric amplifier. The dashed line is the ideal case while the solid lines represent the real case where the periodic variations are due to standing waves in the circulator cable. (b) Gain (color) as a function of signal frequency for different pump frequencies with other bias parameters unchanged. Slight adjustment of the pump frequency allows one to minimize the ripples in the gain profile (black arrow).}
\label{fig:ripple}
\end{figure}

Fig.~\ref{fig:gain} shows the gain profile obtained using our impedance transformer design. The best performance was obtained at $\Omega_d / 2\pi =5.93$ GHz, close to the design frequency of 6 GHz, with a -3 dB bandwidth of 640 MHz (vertical dashed line) and 20 dB gain at the pump frequency. We measured the input impedance of a nominally identical transformer at room temperature (dashed lines of Fig.~\ref{fig:ripple} (a)) and obtained $Z_{\rm aux} = 75 \ \Omega$ and $R = 33 \ \Omega$. We also extracted $R$ from cryogenic measurements using Eq.~\eqref{eq:gmax} and obtained $R=31 \ \Omega$, consistent with our room temperature measurements. Using $R=31 \ \Omega$, the predicted value of bandwidth using Eqs.~(\ref{eq:gain}a) and (\ref{eq:gain}b) is $640$ MHz for $Z_{\text{opt}} = 83 \ \Omega$ which is in excellent agreement with our measured bandwidth. Without the auxiliary resonator the same device (with $R=31 \ \Omega$) would have had 180 MHz of bandwidth, giving us a bandwidth improvement of about a factor of 3.6. The total bandwidth enhancement when compared to the standard case of $R=50 \ \Omega$ is about 5.7.

The noise temperature of the JPA was measured by first calibrating the system noise temperature using a shot noise tunnel junction \cite{Lafe-SNTJ, Lafe-SNTJ-APL} and then switching to the JPA using a cryogenic microwave switch in the same experimental run. The JPA's noise temperature (Fig.~\ref{fig:gain}, red curve) was then extracted by measuring the signal-to-noise ratio improvement obtained using the JPA \cite{supp}. The performance was near quantum-limited ($T_{Q}=\hbar (\Omega_d+\omega) /k_B$) in the full 640 MHz band. The 1-dB compression point is shown in the bottom panel of Fig.~\ref{fig:gain} and has a mean value of about -110 dBm throughout the band.

In practice, the ideal input impedance required  (Eq.~\eqref{eq:zinideal} and dashed lines in Fig.~\ref{fig:ripple}a) is extremely difficult to implement.  A more realistic $Z_\text{in}[\omega]$ is shown in Fig.~\ref{fig:ripple}a (solid lines) where the periodic deviations are primarily due to standing waves in the cable section between the impedance transformer and the circulator (Fig.~\ref{fig:parampLC}(b), henceforth called the circulator cable), and usually lead to ripples in the gain profile.  However,  we could minimize these ripples \cite{supp} by a slight adjustment of the pump frequency. Fig.~\ref{fig:ripple}(b) shows the variation of gain profile as pump frequency is varied while leaving other bias conditions unchanged. The arrow corresponds to the data in Fig.~\ref{fig:gain} where ripples are minimized. This ripple minimization effect can be understood using Eqs.~(\ref{eq:omegakappa}a) and (\ref{eq:omegakappa}b) which show us that variations in $\rm{Im}[Y_\text{in}[\omega]]$ which are symmetric about $\omega=0$ do not affect the gain. By positioning the pump frequency ($\Omega_d$) at the right value with respect to $Y_\text{in}[\omega]$ ,  one can minimize the ripples while not affecting the optimal choice of $\Omega_d=\Omega_\text{aux}$ significantly. The two larger humps at the band edge are probably due to significant deviation of the impedance seen by the JPA from our simplified model. We would like to emphasize that the ripple minimization procedure is robust and we were able to achieve this optimization for several different configurations \cite{supp} with different $Y_\text{in}[\omega]$.

Our scheme thus benefits from having introduced an extra (linear) resonant mode in a judicious manner.  The practical utility of having two non-linear modes was recently discussed in Ref.~\onlinecite{eicher-dimer}, though that design was still subject to the standard gain-bandwidth limitation. It is also interesting to compare our use of two modes to the traveling wave parametric amplifier (TWPA) \cite{TWPA-theory-1} which can be thought of as consisting of a large number of coupled non-linear modes; here, we show that just a single extra linear mode can provide significant advantages.  Further, the phase matching achieved by dispersion engineering the TWPA \cite{TWPA-theory-1} can be compared to our introduction of an imaginary impedance at the signal and idler frequencies to enhance the bandwidth.

In conclusion, we have designed a simple impedance transformer which introduces a positive linear slope in the imaginary part of the impedance seen by the JPA. Our theory explains the significance of the imaginary part of input impedance in enhancing the bandwidth. We demonstrate a device with a nearly flat 20 dB gain over a 3 dB bandwidth of 640 MHz while maintaining near quantum-limited noise performance, in good agreement with theoretical predictions. These characteristics along with a 1-dB compression point of -110 dBm clearly indicates its capability for simultaneous multi-qubit readout \cite{multiplex-martinis}. Our technique is readily adaptable to flux pumped JPAs \cite{flux-driven-Yamamoto}, amplifiers based on the Josephson parametric converter \citep{JPC-amp, JPC-circ}, and flux \cite{nanosquid-Eli} and charge \cite{charge-sensor-paramp} sensors based on parametric amplifiers. It could also be useful as a source of broadband squeezed microwaves \cite{broad-squeeze-2010} for experiments in microwave quantum optics.

This work was supported by the Department of Atomic Energy of Government of India. RV acknowledges funding from the Department of Science and Technology, India via the Ramanujan Fellowship. We also acknowledge the TIFR Nanofabrication facility. AC acknowledges support from the Army Research Office under Grant No. W911NF-14-1-0078.


%

\newpage
\setcounter{figure}{0}
\setcounter{table}{0}
\setcounter{equation}{0}
\onecolumngrid

\global\long\def\theequation{S\arabic{equation}}

\global\long\def\thefigure{S\arabic{figure}}

\vspace{1.0cm}
\begin{center}
{\bf \large Supplementary Material for ``Broadband parametric amplification with impedance engineering: Beyond the gain-bandwidth product"}
\vspace{1.0cm}\\

\end{center}

\section{Device Details}

The JPA device was fabricated on a 275 $\mu$m thick intrinsic silicon chip with standard e-beam lithography and bilayer resist process. The Josephson junctions in the SQUID were made using double-angle evaporation with a 200 \AA{} thick $\mathrm{Al}$ layer at $+30^\circ$, followed by a three minute oxidation in 0.7 Torr of pure oxygen, and a 500 \AA{} thick second layer of $\mathrm{Al}$ at $-30^\circ$. All depositions were done in an e-beam evaporator. From room temperature resistance measurements, the total critical current of the SQUID was estimated to be 2.8 $\mu$A. The capacitors in parallel plate geometry were designed to have an area of $120\times42.5 \ \mu \text{m}^2$ with a 300 \AA{} thick $\mathrm{Al_2O_3}$ sandwiched between two 700 \AA{} thick $\mathrm{Al}$ layers. The Al$_2$O$_3$ and second $\mathrm{Al}$ layer were deposited with the sample stage continuously rotating and a deposition angle of $20^\circ$ and $10^\circ$ respectively to prevent shorting between the two $\mathrm{Al}$ layers. The estimated capacitance was 3.4 pF.
\begin{figure}[b]
\includegraphics[scale=0.8]{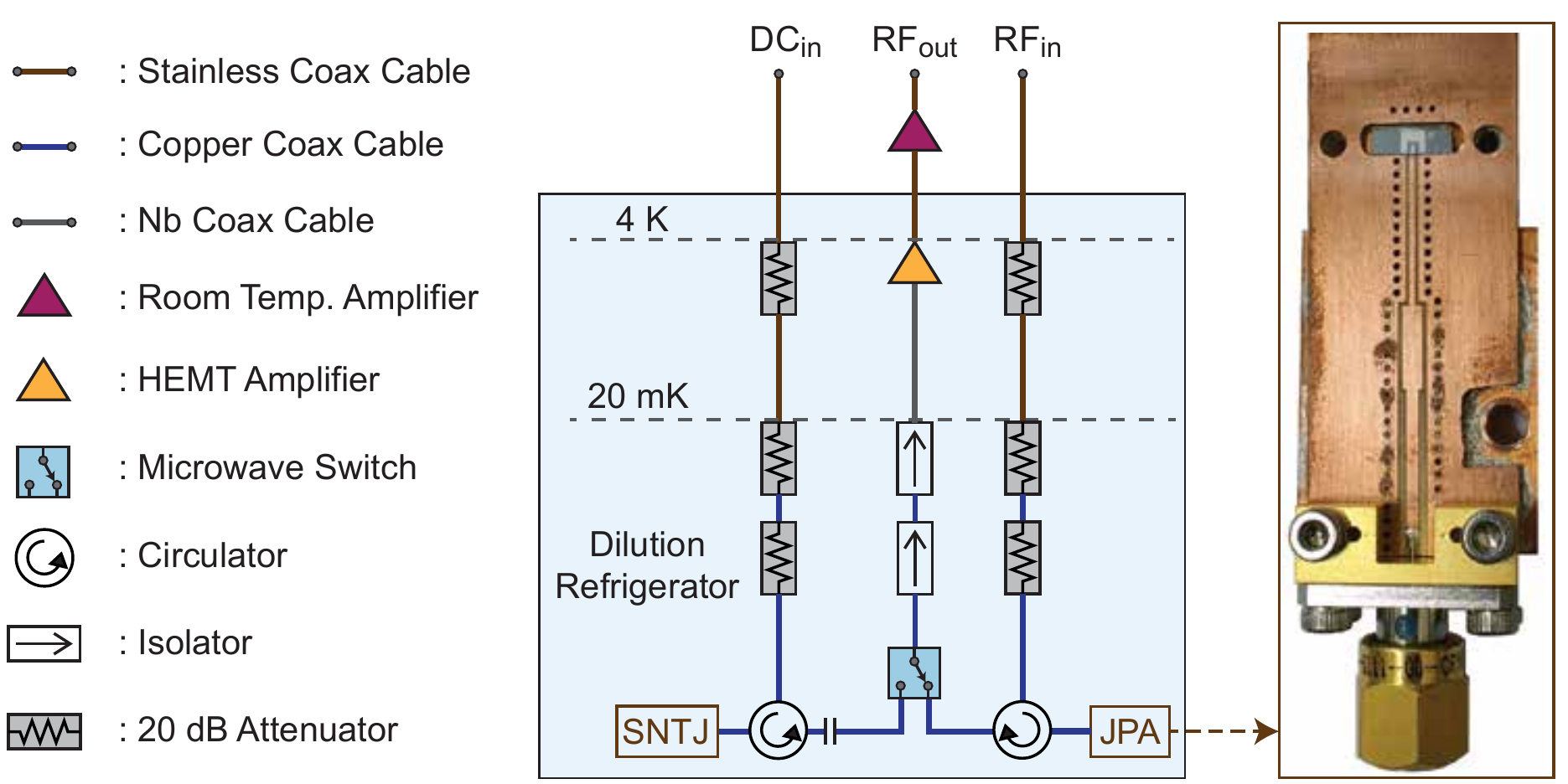}
\caption{Simplified cryogenic microwave setup of our experiment. An optical image of the microwave circuit board showing the CPW impedance transformer and JPA chip is also shown.}
\label{fig:setup}
\end{figure}

\section{Calibration of System Gain and Noise Temperature}
We used a voltage biased tunnel junction as a calibrated source of noise \cite{Lafe-SNTJ-APL-s} to measure the system noise temperature of our amplification chain. The shot noise tunnel junction (SNTJ) is a standard $\mathrm{Al/AlO_x/Al}$ tunnel junction with $50 \ \Omega$ nominal resistance and is embedded in a $50 \ \Omega$ coplanar wave guide transmission line. A strong rare earth magnet is typically used to suppress superconductivity in Aluminum to realize a normal tunnel junction at dilution temperatures. The noise power coupled to a matched load by a voltage biased SNTJ at temperature $T$ and frequency $f$ is given by \cite{Lafe-SNTJ-s}
\begin{equation}
\label{eq:Np}
P_N(f,v)=G_\text{sys}(f) k_BB\left[T_\text{N,sys}(f)+\dfrac{1}{2} \left( \dfrac{eV+hf}{2k_B} \right)\coth\left( \dfrac{eV+hf}{2k_BT} \right)+\dfrac{1}{2}\left( \dfrac{eV-hf}{2k_B} \right)\coth\left( \dfrac{eV-hf}{2k_BT} \right) \right],
\end{equation}
where $G_\text{sys}$ is the gain of the amplification chain, $k_B$ is the Boltzmann constant, $B$ is the bandwidth over which noise is measured, $T_\text{N,sys}$ is the noise temperature of the amplification chain, $h$ is Planck's constant and $V$ is the voltage applied across the tunnel junction. Using the above equation, a measurement of noise power $P_N$ as a function of $V$ allows determination of system temperature $T$, gain $G_\text{sys}$ and most importantly the noise temperature  $T_\text{N,sys}$. Given the system gain, one can extract the input signal power at the plane of the sample allowing one to calibrate the 1-dB compression point as well.

We use the high voltage limit of this equation which results in a linear relation between $P_N$ and $V$ given by
\begin{equation}
\label{eq:Np_lin}
\begin{aligned}
P_N(f,V)=G_\text{sys}(f) k_BB\left( T_\text{N,sys}(f)+\dfrac{eV}{2k_B}\right).
\end{aligned}
\end{equation}
We can then extract $G_\text{sys}$ and $T_\text{N,sys}$ using straight line fits. In a separate experiment, we had ensured that identical results were obtained for $G_\text{sys}$ and $T_\text{N,sys}$ when using Eq.~\eqref{eq:Np} or \eqref{eq:Np_lin}. This allowed us to eliminate the strong magnet since at voltages much greater than the gap, the superconducting tunnel junction behaves like a normal tunnel junction.

\begin{figure}[t]
\includegraphics[width=1\textwidth]{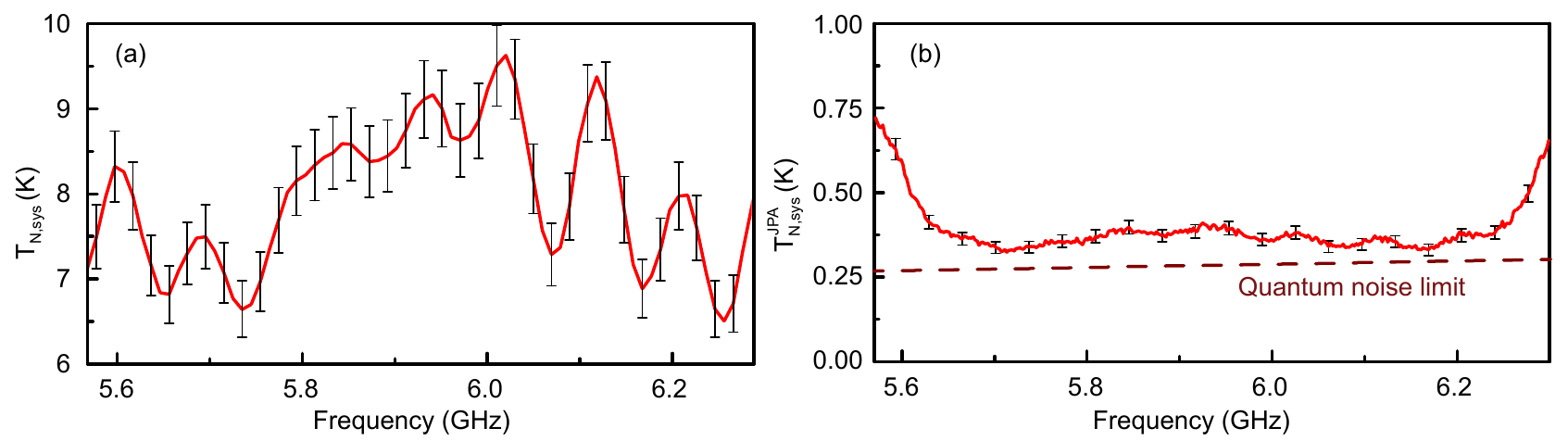}
\caption{Plots of system noise temperature. (a) System noise temperature $T_\text{N,sys}$ of the amplification chain measured using a SNTJ noise source without the JPA. (b) System noise temperature $T_\text{N,sys}^{\rm JPA}$ of the amplification chain with the JPA switched in. The error bars ($\pm 5\%$) reflect the error related to the difference in signal attenuation between the signal paths from the SNTJ and the JPA respectively, to the cryogenic switch.}
\label{fig:tnsys}
\end{figure}

A cryogenic microwave switch was used to route the output of the SNTJ or the JPA to the amplification chain as shown in Fig.~\ref{fig:setup}. With the SNTJ switched in, we measured the system noise temperature $T_\text{N,sys}$ and the result is shown in Fig.~\ref{fig:tnsys}(a). The periodic variations are a result of impedance mismatch primarily in the circulator but the mean value of $T_\text{N,sys}$ was consistent with our HEMT noise temperature after accounting for signal attenuation between the SNTJ and the HEMT.

We tried to keep the signal path from the JPA and the SNTJ to the cryogenic switch identical, so that the measured $T_\text{N,sys}$ (using SNTJ) was still valid with the JPA switched in. However, we estimated that there was a mismatch of about $\epsilon_{att}\sim\pm 0.2$ dB in the transmission between the two signal paths. This directly leads to a fractional error in the estimation of the $T_\text{N,sys}$ of $\pm10^{\epsilon_{att}/10} \sim \pm 5\%$ which is reflected by the error bars in Fig.~\ref{fig:tnsys}(a). Other sources of uncertainty like reduced power coupling due to impedance mismatch of the SNTJ, calibration error in the voltage bias across the SNTJ, and statistical errors in fitting were insignificant in comparison. 

With the JPA switched in, its noise temperature $(T_{N}^{\rm JPA})$ is determined by measuring the signal to noise ratio improvement $\eta_\text{SNR}$ of a probe signal when the JPA is biased to provide 20 dB maximum gain at the pump frequency:
\begin{equation}
T_{N}^{\rm JPA}(f)=T_\text{N,sys}(f)\left[ \eta_\text{SNR}(f)^{-1} - G(f)^{-1}\right]=T_\text{N,sys}^{\rm JPA}(f)-T_\text{N,sys}(f)G(f)^{-1},
\end{equation}
where $G$ is the gain of the JPA. Here $T_\text{N,sys}^{\rm JPA}$ is the system noise temperature with the JPA switched in and is plotted in Fig.~\ref{fig:tnsys}(b). The fractional error in $T_\text{N,sys}^{\rm JPA}$ and $T_{N}^{\rm JPA}$ is the same as that in $T_\text{N,sys}$ which is reflected in the error bars in Fig.~\ref{fig:tnsys}(b) and Fig. 3 in the main text.

\section{Effective two-mode model for the JPA}

Here, we briefly outline why one can use the effective two-mode Hamiltonian to describe the JPA, when it involves only a single physical cavity mode.
The standard approach to a driven JPA is to model it as a driven cavity with a Kerr nonlinearity.  The cavity Hamiltonian in this case (in an interaction picture at the pump frequency) has the 
form\cite{clerk-laflamme-PRA-s}:
\begin{equation}
	\label{eq:jpahamil}
	\hat{H}_A= \omega_{d}' \hat{a}^\dagger\hat{a}  -i \frac{\lambda}{2} (\hat{a}^\dagger\hat{a}^\dagger-h.c.),
\end{equation}
while the equations of motion for the Fourier-transformed cavity mode operators take the form:
\begin{subequations}
\begin{align}
	-i\omega\hat{a}[\omega] & = \left(-i \omega_{d}' - \kappa_0/2 \right) \hat{a}[\omega] 
		- \lambda \hat{a}^\dagger[\omega], \\
	-i\omega\hat{a}^\dagger[\omega] & = \left(i \omega_{d}' - \kappa_0/2 \right) \hat{a}[\omega] 
		- \lambda \hat{a}[\omega],
\end{align}
\end{subequations}
where we have used the following convention for the Fourier-transform
\begin{subequations}
\begin{align}
a[\omega] &=\int a(t)e^{i\omega t}dt, \\
a^\dagger[\omega] &=\int a^\dagger(t)e^{i\omega t}dt.
\end{align}
\end{subequations}
Note that $\hat{a}^\dagger[\omega] = \left( \hat{a}[-\omega] \right)^\dagger$.  For $\omega \neq 0$, the operators $\hat{a}[\omega]$ and $\hat{a}^\dagger[\omega]$ are commuting operators describing independent degrees of freedom (namely photons at a frequency $+\omega$ versus those at $-\omega$). To stress this independence, we can relabel operators (for $\omega > 0$) as:     
\begin{subequations}	
\begin{align}
	\hat{a}[\omega] & \rightarrow \hat{a}_1[\omega], \\
	\hat{a}[-\omega] & \rightarrow \hat{a}_2[-\omega]. 
\end{align}
\end{subequations}
The above equations of motion could then be equivalently derived from the starting Hamiltonian given in Eq.~(1) of the main text.  When we use the JPA as a phase-preserving amplifier, it is convenient to use this effective two-mode description, as then one has a direct equivalence to a standard two-mode non-degenerate parametric amplifier.  While this description is convenient, we stress that identical results would be obtained if one stuck to a single mode description.

\section{Susceptibility matrix from classical equations of motion}
\label{sec:classicaleqns}

\begin{figure}[h]
\includegraphics[width=0.5\textwidth]{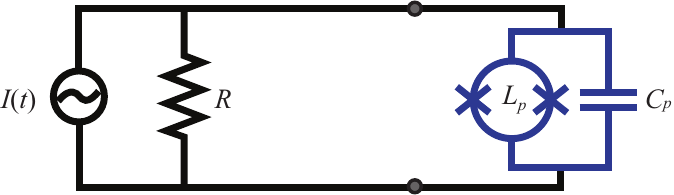} 
\caption{Schematic of a JPA coupled to a frequency independent environment.}
\label{fig:JPAdrive}
\end{figure}

In this section, we will show how the susceptibility matrix $(\chi(\omega))$ can be obtained from the classical equations of motion of a driven non-linear oscillator. Fig.~\ref{fig:JPAdrive} depicts a JPA consisting of a SQUID loop with a total critical current $I_0$ shunted by a capacitor $C_p$. The JPA is coupled to an environment of frequency independent shunt impedance $R$. The equation of motion of the gauge-invariant phase $\delta(t)$ across the Josephson junction is then given by
\begin{equation}
\label{eq:deltaeq}
C_p\varphi_0\dfrac{d^2\delta(t)}{dt^2}+\dfrac{\varphi_0}{R}\dfrac{d\delta(t)}{dt}+I_0\sin\left( \delta(t) \right)=I(t),
\end{equation}
where $\varphi_0=\hbar/(2e)$ is the reduced flux quantum and $I(t)$ is the current applied from an external source. $I(t)$ in general consists of two terms: a drive $I_d(t)$ to bias and a signal $I_s(t)$ to be amplified. Keeping only the first non-linear term in the expansion of $\sin\left( \delta(t)\right)$, Eq.~\eqref{eq:deltaeq} becomes
\begin{equation}
\label{eq:deltaeq3}
\dfrac{d^2\delta(t)}{dt^2}+\kappa_0\dfrac{d\delta(t)}{dt}+\Omega_p^2\left( \delta(t)-\dfrac{\delta(t)^3}{6} \right)=\dfrac{I_d(t)}{C_p\varphi_0}+\dfrac{I_s(t)}{C_p\varphi_0}=A_d\cos{\Omega_d t}+A_s(t),
\end{equation}
where $\kappa_0=1/(RC_p)$ is the damping rate of the oscillator amplitude and $\Omega_p=\left(I_0/(C_p\varphi_0)\right)^{1/2}$ is the plasma frequency of the JPA. The steady state dynamics $(\delta_d(t))$ is controlled by the strong drive $A_d\cos{\Omega_d t}$ and is obtained by solving Eq.~\eqref{eq:deltaeq3} while setting $A_s(t)$ to zero. We write this solution as
\begin{equation}
\label{eq:delta}
\delta_d(t)=\delta_d^\text{max} \cos{\left(\Omega_dt-(\theta+\pi/4)\right)}.
\end{equation}
Next we perform a perturbative expansion around the steady state solution \cite{dykman1994} in order to find the response due to the signal $A_s(t)$ by considering 
\begin{equation}
\label{eq:delta}
\delta(t)=\delta_d(t)+\delta_s(t),
\end{equation}
with the assumption $\delta_s\ll\delta_d$. Substituting $\delta(t)$ into Eq.~\eqref{eq:deltaeq3} and retaining terms only up to first order in $\delta_s(t)$ one obtains the following differential equation
\begin{equation}
\label{eq:deltaseq}
\dfrac{d^2\delta_s(t)}{dt^2}+\kappa_0\dfrac{d\delta_s(t)}{dt}+\Omega_p^2\left( 1-\varepsilon \right)\left( 1-\dfrac{\varepsilon}{1-\varepsilon}\sin{(2\Omega_dt -2\theta)} \right)\delta_s(t)=A_s(t),
\end{equation}
where $\varepsilon=(\delta_d^\text{max})^2/4$. Eq. \eqref{eq:deltaseq} represents a harmonic oscillator driven parametrically at $2\Omega_d$ and its solution can be easily obtained by going to the frequency domain. Taking the Fourier transform of Eq. \eqref{eq:deltaseq}, applying rotating wave approximation, and moving to a  frequency reference centered at $\Omega_d$ one finds
\begin{equation}
\label{eq:fspace00}
\left( -i\omega +i\omega_d'+ \dfrac{\kappa_0}{2} \right) e^{-i\theta}\delta_s[\omega]+\lambda e^{i\theta}\delta_s^\dagger[-\omega]=e^{-i\theta}\dfrac{iA_s[\omega]}{2\Omega_p},
\end{equation}
where $\delta_s[\omega]\equiv\delta_s[\Omega], \ \delta_s^\dagger[-\omega]\equiv\delta_s[\Omega-2\Omega_d], \ A_s[\omega]\equiv A_s[\Omega],$ with $\delta_s[\Omega]$ and $A_s[\Omega]$ being the Fourier transforms of $\delta_s(t)$ and $A_s(t)$ respectively. Here $\lambda=\Omega_p\varepsilon/4=\Omega_p(\delta_d^\text{max})^2/16$ is the effective parametric pump which is controlled by the drive amplitude $A_d$. The other symbols have the same meaning as described in the main text. Defining $\hat{a}_1[\omega] = e^{-i \theta} \delta_s[\omega]$ ,  $\hat{a}_2^\dagger[-\omega] = e^{i \theta} \delta_s^\dagger[- \omega]$ and $\hat{a}_\text{in}[\omega] = i e^{-i \theta}A_s[\omega]/(2 \Omega_p)$, we get,
\begin{equation}
\label{eq:fspace}
\left( -i\omega +i\omega_d'+\dfrac{\kappa_0}{2} \right)\hat{a}_1[\omega]+\lambda \hat{a}_2^\dagger[-\omega]=\hat{a}_\text{in}[\omega].
\end{equation}
Combining  Eq.~\eqref{eq:fspace} along with its Hermitian conjugate, one finds the  susceptibility matrix  $\chi[\omega]$ that relates the intracavity fields $\vec{a}[\omega]=(\hat{a}_1[\omega],\hat{a}_2^\dagger[-\omega])$ to the input fields $\vec{a}_{\rm in}[\omega]$ as $\vec{a}[\omega]=\chi[\omega]\vec{a}_{\rm in}[\omega]$, where
\begin{equation}
\label{eq:genchi}
(\chi[\omega])^{-1}=
-i\begin{pmatrix}
(\omega-\omega_{d}')+i\kappa_0/2& i\lambda \\
i\lambda& (\omega+\omega_{d}')+i\kappa_0/2
\end{pmatrix}.
\end{equation}
This susceptibility is the same as Eq.~(2) in the main text in the absence of any auxiliary resonator so that  $\Sigma_{1(2)}[\omega]=-i \kappa_0/2$. In the most general case, when the environment is frequency dependent, one has to use a frequency dependent self energy to compute the susceptibility matrix as shown in the main text. The self energy is related to the environmental impedance as explained in the next section.

\section{Connection between self energy and the admittance}

\subsection{General linear-response derivation}

In this section, we show how the effects of a general input admittance seen by the JPA can be treated by using a self energy, i.e.~frequency dependent terms which modify the JPA susceptibility defined in Eq.~(2) of the main text.  We stress that as the admittance is associated with a linear system, its treatment via a self energy is exact.  Classically, one can obtain the self energy simply by solving the (linear) equations of motion for the admittance degrees of freedom, and then substituting these into the equations of motion for the JPA degrees of freedom; this is shown explicitly in the next section.  Here, we present a derivation of the self energy based on quantum linear response.  This approach yields additional insights, and is convenient if one is also interested in understanding quantum noise properties of the amplifier.  We stress that the frequency-dependent self energy approach we describe is both a powerful and general way of approaching our system, as it is applicable to an arbitrary input admittance, including cases where several resonances contribute.

We start by working with the single-mode description of the JPA resonator, as described in the previous section, but in the lab frame.  The coupling between the JPA resonator and 
external environment (comprising of the auxiliary circuit plus transmission line) has the form ($\hbar = 1$):
\begin{align}
	\hat{H}_{\rm int} & = 
		\hat{\Phi} \cdot \hat{I}_{\rm env}
		 =
		\sqrt{\frac{\Omega_p L_p}{2}} \left(\hat{a} + \hat{a}^\dagger \right) \cdot \hat{I}_{\rm env}.
\end{align}
Here $\hat{\Phi}$ is the flux operator for the JPA cavity mode, and $\hat{I}_{\rm env}$ is the current operator for the external environment.  As the environment is a linear circuit, it corresponds to a collection of independent bosonic modes (see, e.g., Ref.~\onlinecite{devoret-les-houches}); further, $\hat{I}_{\rm env}$ is linear in the creation and annihilation operators for these modes.  It follows that the retarded self energy $\Sigma_\text{lab}[\omega]$ we want is directly
determined by the retarded correlation function of $\hat{I}_{\rm env}$:
\begin{subequations}
\begin{align}
	\Sigma_{\rm lab}[\omega] &=
		\frac{ \Omega_p L_p}{2} G^R[\omega]  \\
		& = \frac{\Omega_p L_p}{2} \left(-i \int_0^{\infty} dt \,  \langle \hat{I}_{\rm env}(t) \hat{I}_{\rm env}(0) \rangle e^{i \omega t} \right) \\
		& = \frac{ \Omega_p L_p}{2} \left(-i \omega \tilde{Y}_{\rm in,lab}[\omega] \right).
\end{align}
\end{subequations}
In the last line, we have used the Kubo formula to relate the required correlation function to the admittance of the environment $\tilde{Y}_{\rm in,lab}[\omega]$ as seen by the JPA. 
The tilde here indicates that we are using the standard physicist convention for the Fourier transform in writing the admittance.  The frequency dependent admittance using the standard engineering convention $Y_{\rm in, lab}[\omega]$ is given by  $Y_{\rm in, lab}[\omega] = \tilde{Y}_{\rm in, lab}[\omega]^*$.  We use this in what follows.

Finally, we move to the interaction picture (with respect to the drive frequency $\Omega_d$), and use our two-mode representation (c.f.~Eq.~(1) of the main text).  This lets us identify the self-energies for each of these effective modes.  For the effective signal mode self energy (frequencies above $\Omega_d$), we have:
\begin{subequations}
\begin{align}
	\Sigma_{1}[\omega] 
		&\equiv 
			\Sigma_{\rm lab}[\Omega_d + \omega]
			= -i \frac{ \Omega_p L_p}{2} 
				\left(\Omega_d + \omega \right) Y_{\rm in,lab}^*[\Omega_d + \omega]  \\
		& \simeq 
			-i \frac{ 1 }{2 C_p}  Y_{\rm in,lab}^* [\Omega_d + \omega], 
\end{align}
\end{subequations}
where in the last line, we are using the fact that we will be interested in signal frequencies $\omega$ satisfying $| \omega | \ll \Omega_d$ and $\Omega_d\approx\Omega_p$.  Similarly,
\begin{subequations}
\begin{align}
	\Sigma_2[\omega] & \equiv 
					-\left(\Sigma_{\rm lab}[\Omega_d - \omega]\right)^* \\
		& \simeq 
			-i \frac{ 1 }{2 C_p}  Y_{\rm in,lab} [\Omega_d - \omega].
\end{align}
\end{subequations}

Again, we stress that apart from simplifying to frequencies $| \omega | \ll \Omega_d$, this derivation is exact, for the simple reason that linear response is exact for a linear system (i.e.~the input admittance).  The possible hybridization between resonances in the input admittance and the JPA is fully retained in the self energy approach; we show this explicitly below.

\subsection{Self energy for a single-pole input admittance and connection to normal modes}

In the main part of the text, we focus on the simple case where the input admittance seen by the JPA corresponds to a single damped series LC resonance. The self energy corresponding to Eqs.~(9) of the main text is simply:
\begin{eqnarray}
	\Sigma_1[\omega] = \Sigma_2[\omega] =  \frac{1}{2 \alpha C_p} \frac{1}{\omega + i R / \alpha}.
\end{eqnarray}
The simple pole in this self energy reflects the response of the series LC resonance (whose resonance frequency is shifted to zero in our rotating frame).  As our system in this case is composed of two coupled harmonic oscillators (i.e.~the main JPA resonance, and the extra series LC resonance), we expect to find coupled normal modes.  
We stress that this coupling and corresponding hybridization is fully retained in the self energy approach.  Consider the case where there is no effective parametric driving, and hence $\lambda = 0$ in Eq.~(2) of the main text for the susceptibility matrix $\chi[\omega]$.  The susceptibility has simple poles at frequencies satisfying $\omega - \omega'_d - \Sigma_{1/2}[\omega] = 0$.  Using the form for the self energy above, one easily finds that these poles are at $\omega_{\pm}$, with:
\begin{eqnarray}
	\omega_{\pm} = \frac{\omega_d^{\prime}}{2} \pm \frac{1}{2} 
		 \sqrt{ 
		 	 \left(\omega_d^{\prime} + i \frac{R}{\alpha}    \right)^2 
			+ \frac{2}{\alpha C_p} }  
		- i \frac{R}{2 \alpha}.
\end{eqnarray}
These simply represent the two (damped) normal modes of the system.  One can easily see that the effective splitting of these normal modes scales as $1 / \sqrt{\alpha}$, while their broadening scales as $1 / \alpha$.  They thus become resolved in the limit of large $\alpha$ (i.e.~large $Z_{\rm aux}$).

As discussed in the main text after Eq.~(11), one can use the normal mode basis to get an intuitive picture of behavior of the frequency-dependent gain in Fig.~2.  For large 
$Z_{\rm aux}$, the normal modes are resolved, and the two peaks in the gain curve correspond to contributions from each normal mode.  In contrast, for the optimal value of $Z_{\rm aux}$, the two normal modes are not resolved, and they contribute together to yield a gain profile which is extremely flat.  Qualitatively, this behavior can be understood by considering the leading frequency dependence of $\Sigma_j[\omega]$ near $\omega = 0$.  We stress however that our full quantitative results keep the full frequency dependence of $\Sigma_j[\omega]$, and thus are not based on any assumption of an almost frequency-independent self energy.

While the normal-mode picture provides important intuition, we prefer in the main text to focus the discussion on our general self energy approach (as opposed to explicitly focusing on a system with two resonances).  This lets us understand the general principle behind the flattening out of the gain, something that applies even for input admittances that are more complicated than that of a simple series LC resonance.  We also stress that even in the simple case where the input admittance is a series LC circuit, working in the basis of the dissipation-free (i.e.~$R \rightarrow 0$) system normal modes can be  misleading, as the dissipation is strong in the most interesting regime, and because the dissipation is highly asymmetric (e.g. the series LC is damped, the main JPA cavity has negligible damping). 

\section{Obtaining the self energy from the coupled oscillator model}

As discussed in the previous section, given the frequency dependent admittance of the environment one can compute the response of the parametric amplifier. This technique is quite powerful since it allows one to compute the response for any linear circuit which might be used to modify the environment seen by the JPA. In our experiment, we used a combination of $\lambda/4$ and $\lambda/2$ transformers to introduce a positive linear slope in the imaginary part of the input impedance seen by the JPA. The equivalent circuit model is a simple series LC circuit and the corresponding self energy is given by Eqs.~(9a) and (9b) in the main text. We will now show that one gets identical results by considering the same system as two coupled oscillators with a frequency independent environment. 

The classical equations of motion of the circuit in Fig.~1(a) of the main text can be written as
\begin{subequations}
\begin{align}
\label{eq:deltaeq4}
\dfrac{d^2\delta(t)}{dt^2}+\kappa_0\dfrac{d\delta(t)}{dt}+\Omega_p^2\left( \delta(t)-\dfrac{\delta(t)^3}{6} \right)+\dfrac{\kappa_0}{\Omega_d}\dfrac{d^2Q(t)}{dt^2}+\kappa_0\Omega_d Q(t)=A_d\cos{\Omega_d t}+A_s(t), \\
\label{eq:Qeq}
\dfrac{d^2Q(t)}{dt^2}+\kappa\dfrac{dQ(t)}{dt} + \Omega_d^2 \ Q(t)+\kappa_0\dfrac{d\delta(t)}{dt}=A_d\cos{\Omega_d t}+A_s(t),
\end{align}
\end{subequations}
where $Q(t)  = q(t)(\varphi_0 \Omega_d)/ (\kappa_0 Z_\text{aux})$ with $q(t)$ being the charge on the series capacitor $C_\text{aux}$,  $\kappa=\Omega_d R / Z_\text{aux}$ while the other symbols have been defined earlier. We follow the same procedure as described in section~\ref{sec:classicaleqns} and first define the steady state solutions of $\delta(t)$ and $Q(t)$ with $A_s(t)=0$ as $\delta_d(t)$ and $Q_d(t)$ respectively. 
Next we perform a perturbative expansion around the steady state solution in order to find the response due to the signal $A_s(t)$ by considering 
\begin{subequations}
\label{eq:deltaandQ}
\begin{align}
\delta(t)&=\delta_d(t)+\delta_s(t), \\
Q(t)&=Q_d(t)+Q_s(t),
\end{align}
\end{subequations}
where $\delta_d(t)$ is defined in Eq.~\eqref{eq:delta} with the assumption $\delta_s\ll\delta_d$ and $Q_s(t)\ll Q_d(t)$. Substituting $\delta(t)$ and $Q(t)$  into  Eqs.~\eqref{eq:deltaeq4} and \eqref{eq:Qeq} and retaining terms only up to first order in $\delta_s(t)$ one obtains the following differential equations
\begin{subequations}
\begin{align}
\label{eq:deltasandQs}
\dfrac{d^2\delta_s(t)}{dt^2}+\kappa_0\dfrac{d\delta_s(t)}{dt}+\Omega_p^2\left( 1-\varepsilon \right)\left( 1-\dfrac{\varepsilon}{1-\varepsilon}\sin{(2\Omega_dt -2\theta)} \right)\delta_s(t) +\dfrac{d^2Q_s(t)}{dt^2}+ \Omega_d^2 Q_s(t)  =A_s(t), \\
\label{eq:deltasandQsb}
\dfrac{d^2Q_s(t)}{dt^2}+\kappa\dfrac{dQ_s(t)}{dt} + \Omega_d^2 \ Q_s(t)+\kappa_0\dfrac{d\delta_s(t)}{dt}=A_s(t).
\end{align}
\end{subequations}

As before, we now take the Fourier Transform of the above equations, use the rotating wave approximation and move into a reference frequency frame centered at $\Omega_d$. From Eq.~\eqref{eq:deltasandQsb} we can solve for $Q_s[\omega]$ in terms of $\delta_s[\omega]$, and we get
\begin{equation}
\label{eq:Qsw}
Q_s[\omega]= \dfrac{1}{(-i\omega + \kappa/2)}\left( \dfrac{i}{2\Omega_d}A_s[\omega] - \dfrac{\kappa_0}{2}\delta_s[\omega] \right).
\end{equation}
Substituting $Q_s[\omega]$ in the Fourier Transform of Eq.~\eqref{eq:deltasandQs}, we get
\begin{equation}
\label{eq:fspace0}
\left( -i\omega +i\omega_d'+\dfrac{\kappa_0 \kappa}{8}\left( \dfrac{\kappa}{\omega^2 + \kappa^2/4}\right)+\dfrac{\kappa_0 \kappa}{4}\left( \dfrac{i\omega}{\omega^2 + \kappa^2/4}\right) \right) e^{-i\theta}\delta_s[\omega]+\lambda e^{i\theta}\delta_s^\dagger[-\omega]=e^{-i\theta}\dfrac{iA_s[\omega]}{2\Omega_p}\left[1+\dfrac{i\omega}{(-i\omega+\kappa/2)} \right].
\end{equation}
Defining $\hat{a}_1[\omega]=e^{-i\theta}\delta_s[\omega], \ \hat{a}_2^\dagger[-\omega]=e^{i\theta}\delta_s^\dagger[-\omega]$ and $\hat{a}_\text{in}[\omega]$ as the RHS of Eq.~\eqref{eq:fspace0}, we get,
\begin{equation}
\label{eq:fspace1}
\left( -i\omega +i\omega_d'+\dfrac{\kappa_0 \kappa}{8}\left( \dfrac{\kappa}{\omega^2 + \kappa^2/4}\right)+\dfrac{\kappa_0 \kappa}{4}\left( \dfrac{i\omega}{\omega^2 + \kappa^2/4}\right) \right)\hat{a}_1[\omega]+\lambda \hat{a}_2^\dagger[-\omega]=\hat{a}_\text{in}[\omega].
\end{equation}
Comparing Eq.~\eqref{eq:fspace1} and its Hermitian conjugate with Eq.~(2) in the main text, and substituting for $\kappa_0=1/(RC_p)$ and $\kappa=2R/\alpha$, we obtain,
\begin{subequations}
\label{eq:deltakappa1}
\begin{align}
\Delta_{1(2)}[\omega]&=\dfrac{\kappa_0 \kappa}{4}\left( \dfrac{\omega}{\omega^2 + \kappa^2/4}\right) =\dfrac{1}{2 \alpha C_p}\left(\dfrac{\omega}{\omega^2 + R^2/\alpha^2}\right), \\
{\kappa}_{1(2)}[\omega]&=\dfrac{\kappa_0 \kappa}{4}\left( \dfrac{\kappa}{\omega^2 + \kappa^2/4}\right)=\dfrac{R}{\alpha^2 C_p}\left(\dfrac{1}{\omega^2 + R^2/\alpha^2}\right),
\end{align}
\end{subequations}
which are identical to Eqs.~(9a) and (9b) in the main text obtained using the self energy approach.

We would again like to emphasize that the self energy approach is more general and a lot more powerful as it allows us to compute the parametric amplifier response for any linear auxiliary circuit as long as we can calculate the frequency dependent input admittance. The method used in this section will not be amenable for more complex impedance transforming circuits.

\section{Impedance engineering and ripple cancellation}
In what follows, we will write the input admittance as
\begin{subequations}
\label{eq:genZin}
\begin{align}
			\left( Y_{\rm in, lab}[\Omega_d + \omega] \right)^{-1} \equiv \left( Y_{\rm in}[ \omega] \right)^{-1} 
			& = Z_\text{in}[\omega] \\
			& = R_\text{in}[\omega]+iX_\text{in}[\omega]+i\alpha(\omega-\omega_\text{off}),
\end{align}
\end{subequations}
where $R_\text{in}[\omega]$ and $X_\text{in}[\omega]$ are the real and imaginary parts of the environmental impedance (without the auxiliary circuit) which we nominally assume to be $50 \ \Omega$. However, even small impedance mismatches in various components of the measurement chain lead to a frequency dependent complex input impedance. The second imaginary term is due to the auxiliary resonator but it now allows for the more general case where the resonant frequency of the auxiliary circuit can be shifted by $\omega_\text{off}$ with respect to the pump frequency of the JPA. The susceptibility matrix (Eq.~(2) of main text) can be written as
\begin{equation}
(\chi[\omega])^{-1}=
-i\begin{pmatrix}
(\omega-\omega_d')-\Sigma_1[\omega]& i\lambda \\
i\lambda& (\omega+\omega_d') -\Sigma_2[\omega]
\end{pmatrix},
\end{equation}
with $\Sigma_{1(2)}[\omega]=\Delta_{1(2)}[\omega]-i{\kappa}_{1(2)}[\omega]/2$.  The effective ``photon potentials" $\Delta_j[\omega]$ and induced damping rates $\kappa_j[\omega]$ of the JPA are directly determined by the input admittance $Y_\text{in}[\omega] = 1/Z_\text{in}[\omega]$ as 
\begin{subequations}
\begin{align}
\Delta_1[\omega] &= -\text{Im}\left[\dfrac{Y_\text{in}[\omega]}{2C_p}\right], \ \ \Delta_2[\omega] = \text{Im}\left[\dfrac{Y_\text{in}[-\omega]}{2C_p}\right], \label{eq:deltadef} \\
\kappa_1[\omega] &= \ \ \text{Re}\left[\dfrac{Y_\text{in}[\omega]}{C_p}\right], \ \ \kappa_2[\omega] = \ \text{Re}\left[\dfrac{Y_\text{in}[-\omega]}{C_p}\right]. \label{eq:kappadef} 
\end{align}
\end{subequations}

The reflection coefficient is given by
\begin{equation}
\label{eq:rw}
{r}[\omega]= 1-\kappa_1[\omega] \chi_{11}[\omega]= 1-\kappa_1[\omega]\dfrac{-i(\omega+\omega_d'-\Delta_2[\omega])+\kappa_2[\omega]/2}{(-i\tilde\omega[\omega]+\kappa_-[\omega]/2)(-i\tilde\omega[\omega]+\kappa_+[\omega]/2)},
\end{equation}
where
\begin{subequations}
\label{eq:defs}
\begin{align}
\tilde\omega[\omega] &=\omega-\Delta[\omega], \\
 \kappa_\pm[\omega] &=\kappa[\omega]\pm2\sqrt{(\lambda^2-\omega_d'^2)+\epsilon[\omega]},\\
\epsilon[\omega] &= \omega'_d\delta[\omega]-\delta[\omega]^2/4,\\
\delta[\omega] &=(\Delta_2[\omega]-\Delta_1[\omega])+i({\kappa}_1[\omega]-{\kappa}_2[\omega])/2,
\end{align}
\end{subequations}
with 
\begin{subequations}
\label{eq:defs1}
\begin{align}
\Delta[\omega] &= \dfrac{\Delta_1[\omega]+\Delta_2[\omega]}{2} =\text{Im}\left[\dfrac{Y_\text{in}[-\omega]-Y_\text{in}[\omega]}{4C_p}\right], \\
 \kappa[\omega] &=  \dfrac{\kappa_1[\omega]+\kappa_2[\omega]}{2} \ = \text{Re}\left[\dfrac{Y_\text{in}[-\omega]+Y_\text{in}[\omega]}{2C_p}\right] .
\end{align}
\end{subequations}
We have introduced $\Delta[\omega]$ and $\kappa[\omega]$ to capture the effects of the symmetry of $Y_\text{in}[\omega]$ around $\omega=0$. The term $(-i{\tilde\omega}[\omega]+{\kappa}_-[\omega]/2)$ in Eq.~(\ref{eq:rw}) predominantly determines the gain and bandwidth of the JPA and that term depends on $\Delta[\omega]$ and $\kappa[\omega]$ given in Eq.~\eqref{eq:defs1}. Note that $\Delta[\omega]$ and $\kappa[\omega]$ control the impedance engineering effect i.e. they will be constants if the input impedance was a frequency independent $50 \ \Omega$. Eq.~(\ref{eq:defs1}) implies that symmetric variations in $\text{Im}[Y_\text{in}[\omega]]$ around $\omega=0$ will leave $\Delta[\omega]$ unchanged while anti-symmetric variations in $\text{Re}[Y_\text{in}[\omega]]$ leave $\kappa[\omega]$ unchanged. As discussed in the main text, the last term in Eq.~(\ref{eq:genZin}b) which is due to the auxiliary circuit leads to $Y_\text{in}[-\omega] = Y_\text{in}[\omega]^*$ when $\omega_\text{off}=0$. For small $\omega$, this results in $\rm{Im}Y_\text{in}[\omega]$ being anti-symmetric about $\omega$ and hence the gain profile is modified significantly.

\begin{figure}[t]
\includegraphics[width=0.5\textwidth]{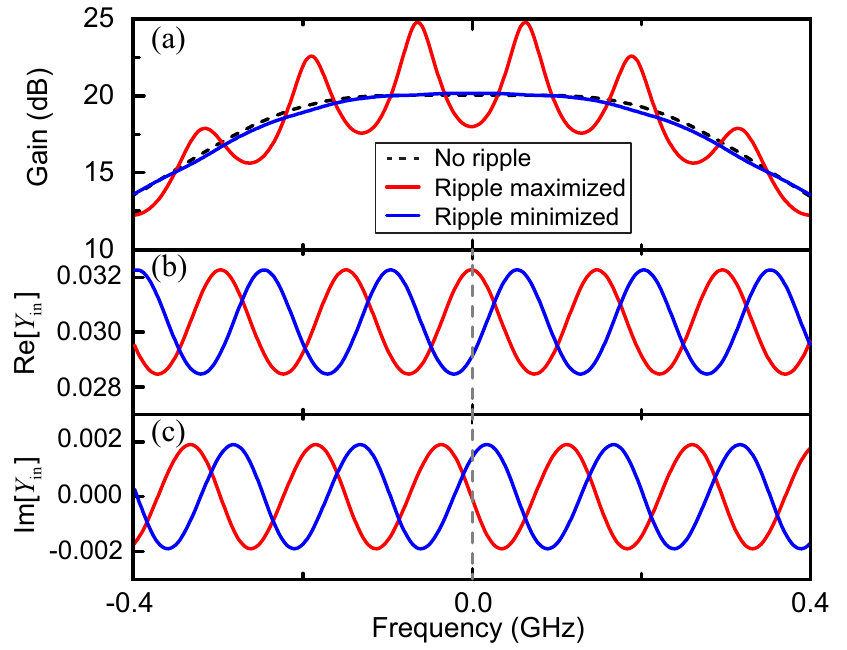}
\caption{Ripple minimization in the gain profile. (a) Gain (theory) with the ripple maximized (red) and minimized (blue) by adjusting the admittance ($Y_\text{in}$) profile shown in (b) and (c). The periodic variations in $Y_\text{in}$ are used to model the effect of the circulator cable and a small impedance mismatch at the circulator. Note that the ripple in the gain is maximized (red curves) when $\text{Re}[Y_\text{in}[\omega]]$ is symmetric and $\text{Im}[Y_\text{in}[\omega]]$  is anti-symmetric about $\omega=0$. By tuning the symmetry of $\text{Re}[Y_\text{in}[\omega]]$ and $\text{Im}[Y_\text{in}[\omega]]$ one can minimize the ripple in the gain (blue curves). The black dashed line in (a) is a reference gain curve with no circulator cable and hence no periodic variations in $\text{Re}[Y_\text{in}[\omega]]$. $Y_\text{in}$ is plotted with $\alpha=0$ for clarity.}
\label{fig:Grip}
\vspace{-0.1cm}
\end{figure}

As mentioned earlier, standing waves resulting from the finite circulator cable length and impedance mismatch at the circulator, lead to a frequency dependence in $R_\text{in}[\omega]$ and $X_\text{in}[\omega]$. Use of a circulator cable is practically unavoidable because one would like to protect the JPA from the magnetic field of the circulator by keeping it inside a magnetic shield and away from the circulator. As shown in Fig.~\ref{fig:Grip}(b),(c), this introduces additional periodic oscillations in $Y_\text{in}[\omega]$ where the period is set by the length of the circulator cable and magnitude by the impedance mismatch. This results in the ripple in the gain profile as shown in the red curve of Fig.~\ref{fig:Grip}(a).

Now if one could adjust the periodic pattern in $Y_\text{in}[\omega]$ in such a way that $\text{Im}[Y_\text{in}[\omega]]$ is symmetric while $\text{Re}[Y_\text{in}[\omega]]$ is anti-symmetric about $\omega=0$, then the ripples in the gain profile can be minimized as shown in the blue curve in Fig.~\ref{fig:Grip}(a). The black dashed curve is the gain profile with no circulator cable shown for reference. This adjustment can be done either by changing the circulator cable length or by  slightly adjusting the pump frequency. However, one cannot completely eliminate these ripples because of the small contributions from $\epsilon[\omega]$ in Eq.~(\ref{eq:defs}b) which actually gets enhanced when one performs the adjustment explained above. Fig.~\ref{fig:Grip}(b) and \ref{fig:Grip}(c) show the real and imaginary parts of $Y_\text{in}[\omega]$ (with $\alpha=0$ for clarity) when the ripple is maximized (red curves) and minimized (blue curves). In practice, the variations in $Y_\text{in}[\omega]$ are not simply sinusoidal but can have an overall frequency dependence which also limits the effectiveness of this trick.

\begin{figure}[t]
\includegraphics[width=0.5\textwidth]{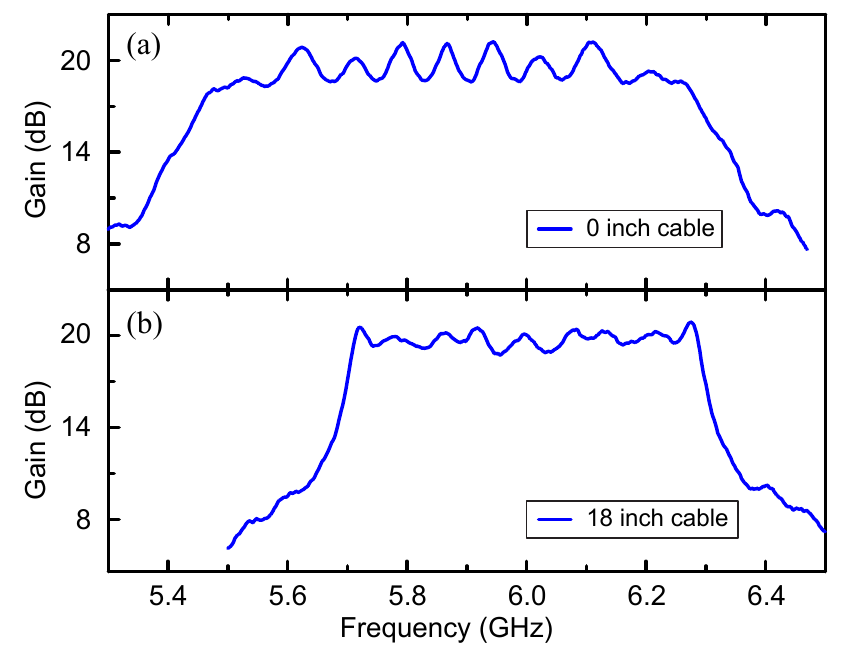}
\caption{Experimental gain profile with the JPA connected directly to the circulator is shown in (a) while (b) shows data with an 18 inch circulator cable. We observe different bandwidths because of differences in $\text{Re}[Y_\text{in}[0]]$. The gain profile in (b) has less ripples than that in (a). We suspect that this is due to the input impedance of the circulator not being symmetric about $\omega=0$ which makes the ripple minimization procedure less effective. Even though the longer cable in (b) adds more ripples to $Y_\text{in}$, it is probably more uniform around $\omega=0$ and hence better suited for ripple minimization.}
\label{fig:Gcomp}
\end{figure}

Since we couldn't tune the circulator cable length in-situ in our experiments, we changed the pump frequency in small steps to achieve the ripple cancellation. Note that this leads to $\omega_\text{off} \neq 0$ and in general affects the gain profile. However, we checked from our theory that provided $\omega_\text{off}$ is small compared to the enhanced bandwidth, it doesn't affect the gain and bandwidth significantly. In order to test the effect of the circulator cable length, we performed different experimental runs with cable sections varying in length. In Fig.~\ref{fig:Gcomp}(a) the JPA was directly mounted (with an SMA male-male barrel) on the circulator showing 830 MHz bandwidth while in Fig.~\ref{fig:Gcomp}(b) we obtained 600 MHz bandwidth with an 18 inch cable section which is three times longer than the cable used to obtain the data in the main text. The large variation in bandwidth can be attributed to different quality factors set by different Re$[Y_\text{in}[0]]$ in these two cases. The effectiveness of ripple minimization also varied between the two samples. Even though the data in Fig.~\ref{fig:Gcomp}(a) shows more bandwidth than the one presented in the main text, the amplifier was significantly noisier. We believe that this is due to the quality factor being too low in this configuration, leading to the amplifier becoming unstable. 

In practice, we don't know the exact value of $\Omega_\text{aux}$ and we tune $\Omega_d$ (around the expected $\Omega_\text{aux}$) and $\Omega_p$ (by flux) till we get $\mathcal{G}_\text{max} \approx 20 \ \text{dB}$ and the largest bandwidth while ensuring minimal ripples in the gain profile. The ripple minimization procedure is robust and we were able to achieve this optimization for several different configurations with different $Y_\text{in}[\omega]$) as shown in Fig.~\ref{fig:Gcomp}.

\section{Calculation of $Z_\text{opt}$}

In order to compute $Z_\text{opt}$, we reproduce Eqs.~(9a,b) from the main text below:
\begin{subequations}
\label{eq:deltakappa-s}
\begin{align}
\Delta_{1(2)}[\omega]&=\dfrac{1}{2 \alpha C_p}\left(\dfrac{\omega}{\omega^2 + R^2/\alpha^2}\right), \\
{\kappa}_{1(2)}[\omega]&=\dfrac{R}{\alpha^2 C_p}\left(\dfrac{1}{\omega^2 + R^2/\alpha^2}\right).
\end{align}
\end{subequations}
Here, $\alpha = 2 Z_\text{aux}/\Omega_d$ is the tuning parameter which can be used to cancel the $\omega^2$ dependence of $\tilde{\mathcal{G}}[\omega]$ for enhancing bandwidth. At the maximum gain point with $\lambda=\lambda_\text{max}=\kappa_0\sqrt{1+3\beta^2}/\sqrt{12}$, the optimal value of $Z_\text{aux}$ is obtained by setting the coefficient of $\omega^2$ in $\tilde{\mathcal{G}}[\omega]$ to zero and we get
\begin{equation}
\label{eq:optz0}
Z_\text{opt}=\eta \dfrac{R^2}{Z_p},  \ \ \ \eta=\dfrac{\Omega_d}{\Omega_p}\dfrac{1}{1+\sqrt2 \sqrt{1+3\beta^2-2\beta\sqrt{1+3\beta^2}}}.
\end{equation}

\section{Impedance transformer design}

The input impedance seen looking towards a load $Z_L$ through a transmission line section of impedance $Z_0$ and length (in units of wavenumber) $\bar{\nu}$ for a design frequency $\Omega_0$ is given by \cite{pozar-book}
\begin{equation}
\label{eq:Zin}
Z_\text{in}[\omega]=Z_0 \dfrac{Z_L+i Z_0 \tan\left(2\pi \bar{\nu} (1+\omega/\Omega_0)\right)}{Z_0+i Z_L \tan(2\pi \bar{\nu} (1+\omega/\Omega_0))}.
\end{equation}

Eq.~\eqref{eq:Zin} shows that any $\lambda/2$ $(\bar{\nu}=1/2)$ or $\lambda/4$ $(\bar{\nu}=1/4)$ transformer provides a positive linear slope in the imaginary component of $Z_\text{in}$ close to its design frequency. For small values of $\omega$, we can model this slope with a series LC circuit with characteristic impedance $Z_\text{aux}$. In case of a $\lambda/2$ transformer of impedance $Z_{\lambda/2}$, the characteristic impedance is given by
\begin{equation}
Z_{\text{aux},\lambda/2}=Z_{\lambda/2}\left( 1-\dfrac{Z_L^2}{Z_{\lambda/2}^2}\right)\dfrac{\pi}{2}.
\end{equation}
Similarly, for a $\lambda/4$ transformer of impedance $Z_{\lambda/4}$, 
\begin{equation}
Z_{\text{aux},\lambda/4}=Z_{\lambda/4}\left( 1-\dfrac{Z_{\lambda/4}^2}{Z_L^2}\right)\dfrac{\pi}{4}.
\end{equation}
Note that the $\lambda/4$ section also transforms the real part of the impedance. However the frequency dependence in the real part over the frequency range of interest is quite small and can be ignored for computing the bandwidth.
\begin{figure}[t]
\includegraphics[scale=1.2]{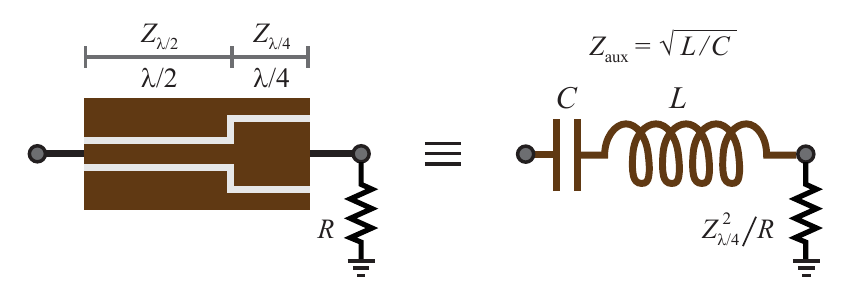}
\caption{Modeling the impedance transformer as a series LC circuit.}
\label{fig:ZLC}
\end{figure}
Since we used a $\lambda/4$ transformer followed by a  $\lambda/2$ transformer, the characteristic impedance of the effective auxiliary resonator can be written as
\begin{equation}
Z_\text{aux}=\left[ Z_{\lambda/4}\left( 1-\dfrac{Z_{\lambda/4}^2}{R^2}\right)+ 2Z_{\lambda/2}\left( 1-\dfrac{Z_{\lambda/4}^4}{R^2 Z_{\lambda/2}^2}\right)\right] \dfrac{\pi}{4},
\end{equation}
where we have substituted $Z_L = R$, the characteristic impedance of the measurement chain. We designed our impedance transformer to have $Z_{\lambda/4}=40 \ \Omega$ and $Z_{\lambda/2}=58 \ \Omega$ resulting in $Z_\text{aux}=75 \ \Omega$ with $R=50 \ \Omega$.


%

\end{document}